\documentclass[%
 aip,
 amsmath,amssymb,
preprint,%
]{revtex4-1}

\usepackage{graphicx}
\usepackage{dcolumn}
\usepackage{bm}
\usepackage[utf8]{inputenc}
\usepackage[T1]{fontenc}
\usepackage{mathptmx}
\usepackage{xcolor}
\usepackage{url}
\usepackage{hyperref}
\usepackage[normalem]{ulem}
\usepackage{subfigure}

\hypersetup{
 colorlinks = true,
 allbordercolors = {white},
 allcolors = {blue},
 }
\usepackage[symbol]{footmisc}

\begin{document}

\title{Bifurcation sequence of two-dimensional Taylor-Green vortex via vortex interactions: Evolution of energy spectrum}

\author{Tapan K. Sengupta\footnote{Corresponding author.}}
\email{tksengupta@iitism.ac.in}
\affiliation{Department of Mechanical Engineering, IIT (ISM) Dhanbad, Jharkhand-826 004, India}

\author{Ankan Sarkar}
\affiliation{Department of Mechanical Engineering, IIT (ISM) Dhanbad, Jharkhand-826 004, India}

\author{Bhavna Joshi}
\affiliation{Department of Mechanical Engineering, IIT (ISM) Dhanbad, Jharkhand-826 004, India}

\author{Prasannabalaji Sundaram}
\affiliation{CERFACS, Toulouse, France}

\author{V.K. Suman}
\affiliation{Department of Aerospace Engineering, IIT Kanpur, U.P.-208 016, India}


 


\date{\today}

\begin{abstract}
The vorticity dynamics of the two-dimensional (2D) Taylor-Green vortex (TGV) problem is investigated in its multi-cellular configuration by solving the incompressible Navier-Stokes equation for long time intervals using a pseudo-spectral method. This helps follow the vorticity dynamics of periodic free shear layer flows by solving an extremely accurate algorithm to explain vortex interactions that lead to vortex stripping (forward cascade), merger, and reconnection (inverse cascade) during various stages of evolution of periodic arrangements of a large number of TGV vortical cells. This latter aspect has been adopted so as not to be affected by the periodicity constraints of a single periodic cell and the various imposed symmetries that attenuate disturbance growth. The analytic solution of the TGV provides the initial condition and the spatially accurate Fourier spectral method enables one to track the first instability of the initial doubly periodic vortices. Despite a plethora of studies following the primary instability to relate it with transition to turbulence and the subsequent decay of turbulence in the literature, the topic of bifurcation sequence for periodic TGV is rare, and that is one of the main aims of the present research. Instead of restricting one's attention on a single periodic TGV cell, here it is purposely reported for multiple cells of the TGV in both directions, without invoking any asymmetries extraneously. For such an ensemble, one can study various vortical interactions giving rise to atypical energy spectra, a topic that has also been seldom addressed to distinguish between successive instabilities that can upon a conjecture, lead to transition and subsequent relaminarization, versus the bifurcation sequences leading from one equilibrium state to subsequent ones. The present study shows the dominance of the latter for 2D TGV at post-critical Reynolds number.
\end{abstract}

\maketitle

\\
\section{Introduction}
Taylor-Green vortex problem\cite{Taylor_Green} occupies an important role in understanding theoretical and numerical aspects of free shear layer vortical flows with enforced periodicity. This is due to the existence of an analytical solution that is periodic in 2D space, allowing one to use Fourier basis functions for spatial discretization \cite{Canuto_etal, Gottleib_Orszag, Brachet_etal1983, Brachet1991, Brachet_etal1992} in pseudo-spectral methods for enhanced accuracy.


Existence of an analytical solution enabled Taylor and Green to study the TGV problem by using a perturbation series in time \cite{Taylor_Green} in an apparent attempt to explain transition to turbulence. Goldstein also extended this perturbation series analysis by expanding it in terms of the Reynolds number ($Re$) - which can be taken as the inverse of the kinematic viscosity ($\nu$). These analyses displayed singularity in time and $Re$, prompting researchers to interpret such singularities as the harbinger of turbulence. Many researchers thereafter numerically investigated the TGV problem by solving the Euler equation \cite{MOF, Brachet_etal1983, Brachet_etal1992}. Brachet {\it et al.} also \cite{Brachet_etal1983} investigated the three-dimensional (3D) viscous flow problem for the generation of small-scale structures by vortex stretching in the resulting turbulence. It was noted that "the inviscid dynamics are strongly influenced by symmetries which confine the flow to an impermeable box with stress-free boundaries" and the resultant evidence of vortex stretching made the authors suggest that "more violent vortex stretching takes place at later times" for which more sophisticated analysis was required. The reported viscous analysis displayed "roll up" of vortex sheets created by inviscid mechanism to suffer instabilities and make the flow chaotic to turbulence state displaying small scales of the flow to be isotropic for high $Re$.

The TGV problem in either 2D or 3D form allows one to study the primary instability of an unsteady equilibrium flow, as has been successfully reported in \cite{Sengupta_etal2DTGV, Gau_Hattori, Sharma_etal2020}. These studies being solely interested in studying the primary instability, the governing Navier-Stokes equation have been solved for a limited extent of time. Brachet \cite{Brachet1991} reported a DNS of 3D TGV using a $864^3$ grid and reaffirmed the early observation of the presence of the first two stages of vorticity dynamics; namely the inviscid stage of creation of vorticity gradient followed by the creation of small scales of turbulence with the computation time limited up to $t =10$, which corresponds to a time little later than when the energy dissipation reaches its maximum value at $t \approx 9$. In this study, a major observation was relating vorticity dynamics with pressure field in writing down the Poisson equation for the static pressure, with the forcing given by the enstrophy and energy dissipation rate terms appearing with opposite signs. It was noted that the vortices are strongly correlated with low-pressure regions. All the subsequent 3D DNS efforts by Brachet and co-authors did not specifically use any explicit excitation. 

The same approach of using the equilibrium solution of the TGV problem given by Taylor and Green \cite{Taylor_Green} was adopted by Sharma and co-authors \cite{Sharma_Sengupta2019, Sharma_etal2020} without any explicit excitation for 3D TGV problem. Instabilities occured via the growth of background disturbances due to numerical errors with the computations performed using near-spectral accurate compact schemes. The Navier-Stokes equations were solved in vorticity-vector potential formulation, that exactly satisfies the solenoidality condition for the vector potential and vorticity. 

All the 3D TGV studies are restricted to small times for direct numerical simulation (DNS) either by pseudo-spectral or by compact schemes on non-uniform grids. While both provide very high spectral spatial accuracy, the cost of such computing prohibits very long simulation time. This motivates one to use pseudo-spectral method for 2D TGV problem computed over extremely large time interval to follow (a) the vorticity dynamics and (b) seek the existence of other equilibrium state(s); when the problem is solved in a multi-cellular configuration incorporating more than one period of the basic TGV unit. Such a study would enable one to track various vortical interactions like vortex stripping \cite{Mariotti_etal}, merger and reconnection \cite{Saffman} etc. over the protracted period recording the resultant vorticity dynamics. The existence of other equilibrium states is equally interesting in its own right, as there are other instances in fluid dynamics where the primary instability is not followed by other instabilities; instead the unstable primary growth is followed by a nonlinear saturation into another limit cycle. A typical example is the flow past a circular cylinder, where the primary temporal instability of the steady state is followed by a nonlinear saturation which has been shown to be governed by Stuart-Landau-Ekhaus equation \cite{Dyn_JFM, Sengupta21, Sengupta_etalPRE2015} expressing the resultant Hopf bifurcation (see Chapter 6 of Sengupta \cite{IFTT} for details). One of the primary goals of the present work is to explore the existence of bifurcation sequence for 2D TGV problem, and the study of the 2D problem is even more desirable as such flows do not have the presence of the destabilizing vortex stretching mechanism which is always present for the 3D TGV problem.  






It is important to highlight the accuracy of the pseudo-spectral methods in the literature, in the context of the global spectral analysis which is used to calibrate numerical methods \cite{GSA_2023}. This is due to the fact that for flows experiencing physical instabilities it is important to investigate spatio-temporal accuracy of the discretization together that instead of only considering the accuracy of spatial discretization. Pseudo-spectral methods have been increasingly used in recent times, as noted for the DNS of homogeneous isotropic turbulence \cite{Buaria_etal2020}, reporting solution using $12288^3$ periodic 3D grid points. One of the striking issues of such DNS is the use of a forcing based on the original work of Rogallo \cite{Rogallo} which requires hyperviscosity for the suppression of numerical instability \cite{arxiv_DNS2021, TKS_IUTAM_Goa}. The latter arises in those efforts that use a two-stage Runge-Kutta method, while the numerical instability is milder for three-stage Runge-Kutta time integration method. This has been thoroughly investigated with the canonical convection and convection-diffusion equations, and it is shown that the four-stage Runge-Kutta method (RK4) is significantly superior as instability occurs at a significantly higher CFL number \cite{arxiv_DNS2021}. Thus in the reported results here, the pseudo-spectral method is used with the RK4 time integration scheme for long-time integration of the Navier-Stokes equation for the 2D TGV problem. The only issue for the simulation of 2D TGV problem is related to aliasing caused by the convection terms, and can be controlled by choosing correct de-aliasing techniques. For example, using the 3/2-rule of zero-padding \cite{Canuto_etal}, one partially circumvents aliasing error. A correct zero-padding has been proposed with 2-rule \cite{FCFD} that ensures complete removal of aliasing error.

Brachet {\it et al.} \cite{Brachet_etal1988} studied the {\it free decaying turbulence} for the 2D TGV problem, where the {\it turbulence} was initiated by random excitation as the initial condition, i.e. without using the initial condition provided by Taylor and Green \cite{Taylor_Green}. The authors interpreted an inertial range from the energy spectrum exponent changing from -4 to -3. The lower value of the exponent was identified to be associated with an isolated vorticity gradient sheet, as postulated by Saffman \cite{Saffman} while studying the interaction between vortex rings via reconnection. The second exponent (-3) is typical of 2D turbulence, as given by Kraichnan \cite{Kraichnan} and Batchelor \cite{Batchelor}, where also the turbulence is attributed to the enstrophy cascade \cite{Doerring_Gibbon, ETE_2013}. 


We note that there has been a linear instability study using modal and non-modal approaches for the 2D TGV problem \cite{Gau_Hattori}. This linear stability study required enforcing strict symmetry via the boundary conditions for a special steady-state equilibrium flow which considered the initial vortices to be elliptic. The 2D TGV problem has also been investigated in Sengupta {\it et al.} \cite{Sengupta_etal2DTGV} with the specific aim of understanding the primary instability with the help of a disturbance enstrophy transport equation (DETE) \cite{Asengupta_etal2018} developed by the authors from the solution of the Navier-Stokes equation. The high accuracy solution is obtained using the stream function- vorticity formulation with the help of a newly developed non-uniform compact scheme\cite{NUC6}. The DETE has its genesis in the enstrophy transport equation \cite{ETE_2013} derived from the first principle for 2D and 3D incompressible flows. 


The instability of the 2D TGV problem was investigated with (2 $\times$ 2)-vortical cells, with the initial condition given by the equilibrium solution of Taylor and Green, so that there is a single full-saddle point in the center of the computational domain \cite{Sengupta_etal2DTGV}. From the solution of the Navier-Stokes equation, the linearised disturbance enstrophy was obtained as, $\Omega_{ld} = 2 \vec{\omega}_m \cdot \vec{\omega}_d$, with $\omega$ indicating the vorticity and the subscripts $m$ and $d$, indicate the equilibrium and disturbance quantities, respectively. Two possibilities for the instability are indicated: (i) When $\frac{D\Omega_{ld}}{Dt} > 0$ for $\Omega_{ld} >0$ and (ii) when $\frac{D\Omega_{ld}}{Dt} < 0$ for $\Omega_{ld} < 0$. The flow is computed for a relatively shorter time interval, up to the primary instability stage \cite{Sengupta_etal2DTGV}.

The study of 2D TGV by Brachet {\it et al.} \cite{Brachet_etal1988} of decaying turbulence initiated by random excitation raises the following queries: (a) Is the 2D turbulence a consequence of the random excitation for the pseudo-spectral method used by them? (b) What will happen to the vorticity dynamics, if the study is extended to larger domain size with correspondingly larger number vortical elements without any stochastic excitation via initial and/ or boundary condition following the deterministic dynamics of the Navier-Stokes equation? These are the motivations of the present study here using the Fourier-Spectral method with RK4 time integration scheme for which definite global spectral analysis results have been reported only recently \cite{arxiv_DNS2021}. Additionally, the energy spectrum would be tracked for the multi-cellular 2D TGV problem for different heights in the larger computational domain for the ($4 \times 4$)-vortical cells, as compared to that in Sengupta {\it et al.} \cite{Sengupta_etal2DTGV}.

In the present research, the 2D TGV problem is solved with $256 \times 256$ uniformly spaced grid points in a domain given by $0 \leq (x,y) \leq 4\pi$. The main aim is to integrate the Navier-Stokes equation for a very long time, to trace the resultant vorticity field and draw definitive results for the physical mechanisms during the evolution of the vorticity field; specifically, the energy spectrum is reported during the time that takes the flow back to an ordered minimal configuration.  

The paper is formatted in the following manner. In the next section, the equilibrium flow of the 2D TGV problem is described briefly. The computational details are provided in Section III, providing the main numerical parameters. Section IV provides the episodic description of the evolving vorticity dynamics and enstrophy. The vorticity dynamics is described in terms of energy and enstrophy variation in Section V. Further post-processing of the evolving vorticity field is described also in Section VI, by displaying the energy spectrum of flow field at selected vertical locations in the computational domain. The paper closes with a summary and conclusion of the present investigation in section VII.

\section{The Equilibrium 2D Taylor-Green Vortex Flow}
\label{sec:equl}
The time-dependent 2D incompressible Navier-Stokes equations are solved using stream function ($\psi$) and vorticity ($\omega$) formulation. The resultant stream function equation (SFE) and the vorticity transport equation (VTE) are given in Cartesian frame by \cite{Sengupta_etal2DTGV, HACM},
  
\begin{equation}
\label{tgVsfe}
\frac{\partial^2 \psi}{\partial x^2} + \frac{\partial^2 \psi}{\partial y^2} = -\omega
\end{equation}

\begin{equation}
\label{tgVvte}
    \frac{\partial \omega}{\partial t} + u\frac{\partial \omega}{\partial x} + v\frac{\partial \omega}{\partial y} = \frac{1}{Re} \left(\frac{\partial^2 \omega}{\partial x^2} + \frac{\partial^2 \omega}{\partial y^2} \right)
\end{equation}

The stream function and the velocity are related by 
$\vec{V}  = \nabla \times \boldsymbol{{\vec{\psi}}}$, with $\boldsymbol{{\vec{\psi}}} = [0 \ 0 \ \psi]^T$, and the vorticity ($\boldsymbol{{\vec{\omega}}}$) can be expressed in terms of the velocity by: $\boldsymbol{{\vec{\omega}}} = \nabla \times\vec{V}$. The components of velocity vector are obtained from, $u = \frac{\partial \psi}{\partial y}$ and $v = -\frac{\partial \psi}{\partial x}$. The TGV problem is solved in a periodic domain 0 $\leq (x,y) \leq 4\pi$ with the following initial conditions \cite{Taylor_Green}, 
\begin{align}
\psi(x,y,0) &= \sin x\; \sin y, &\omega(x,y,0) &= 2 \sin x\; \sin y
\end{align}

The double-periodic, time-dependent analytical solution can be derived using the following ansatz, 
\begin{equation}
\begin{aligned}
\psi(x,y,t) &= \sin x\; \sin y\; F(t) \\ \omega(x,y,t) &= 2 \sin x\; \sin y\; F(t)
\end{aligned}
\label{eqAn}
\end{equation}
with $F(t)$ obtained by substituting Eq. \eqref{eqAn} in Eq. \eqref{tgVvte} as, 
\begin{equation}
	F(t)= e^{-\frac{2t}{Re}}
\end{equation}

\begin{figure*}
\centering
\includegraphics[width=1\linewidth]{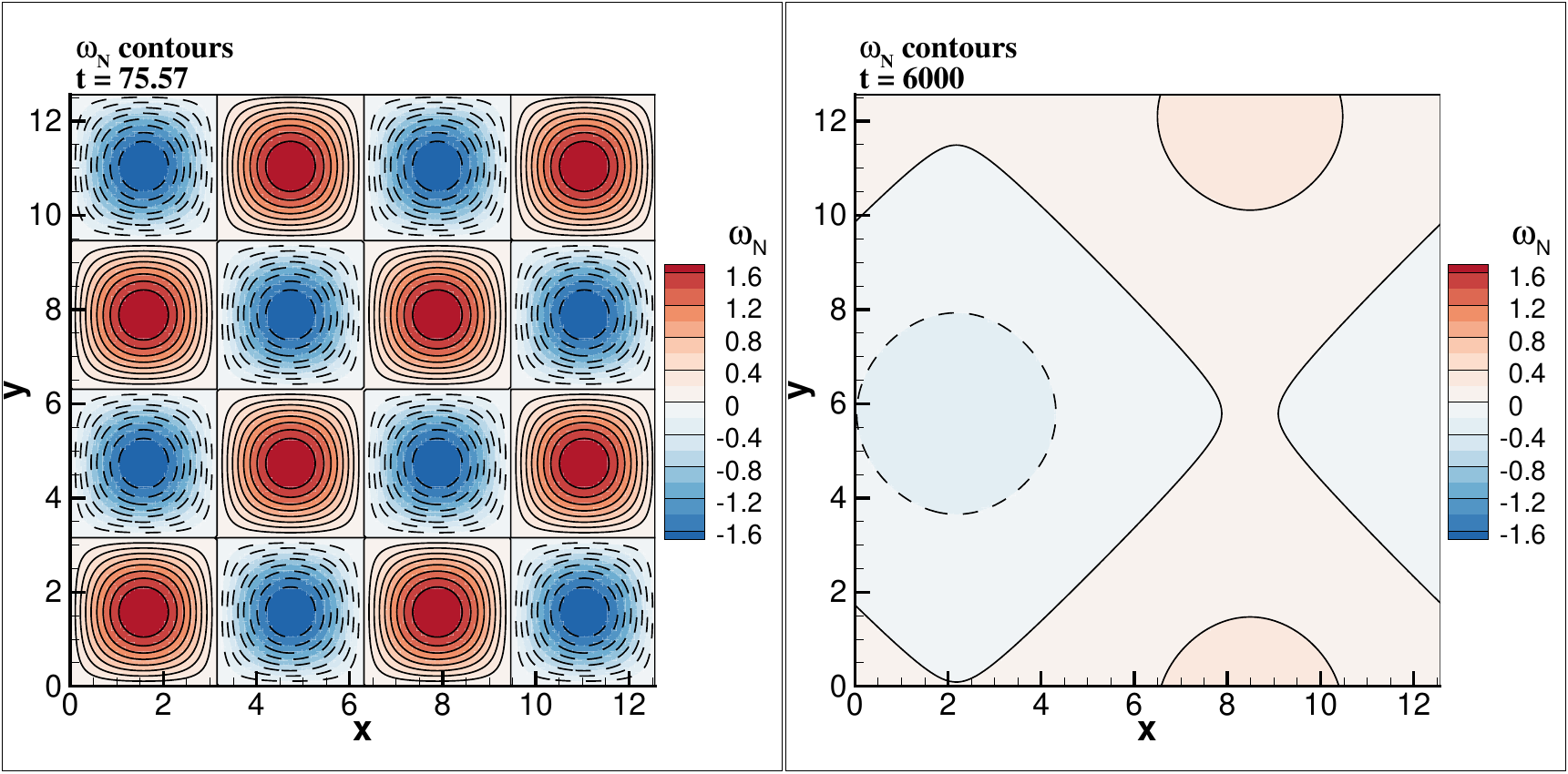}
\caption {The initial analytical solution for the vorticity contours at $t=0$ is shown in the domain $0 \leq (x,y) \leq 4\pi$, with sixteen vortical cells as ordered numerically from 1 to 16, in the left frame. The contours with solid lines are for positive vortices and the dashed lines are for the negative vortices. In the right frame, the computed vorticity contour at $t = 6000$ is shown, as obtained by solving the 2D Navier-Stokes equation by the pseudo-spectral method with RK4 time marching scheme for $Re = 2000$ solved with $(256 \times 256)$ uniform grid.}
\label{Fig1}
\end{figure*}

This is the equilibrium solution for the 2D TGV problem, whose instability to omnipresent numerical disturbances has been reported  \cite{Sengupta_etal2DTGV}. In Fig. \ref{Fig1}, the analytical solution for the initial vorticity contours is shown in the domain $0 \leq (x,y) \leq 4\pi$, with sixteen vortical cells ordered numerically from 1 to 16 in the left frame. The contours with solid (red) lines are for positive vortices and the dashed (blue) lines are for negative vortices. As the governing equation is solved, the vortices will convect, diffuse, and interact among themselves involving the basic processes described in the introduction. The high accuracy afforded by the Fourier spectral method with RK4 time marching scheme allows us to compute the flow indefinitely. To demonstrate this aspect of the present computations, on the right frame of Fig. \ref{Fig1}, the computed vorticity contours shown at $t = 6000$ demonstrate the orderly state of the flow field with two weak vortices surviving. In the following, we describe the vorticity dynamics at the intermediate times. However, before that we highlight the important aspects of the pseudo-spectral method employed here in the next section III.

\section{Pseudo-spectral method for the 2D incompressible Navier-Stokes equations in ($\psi$,$\omega$)-formulation}

The pseudo-spectral method as a powerful and efficient numerical technique can be used to solve partial differential equations (PDEs) in various scientific and engineering fields which are periodic in space. Specifically, in the context of fluid dynamics, it proves to be highly effective for solving the 2D incompressible Navier-Stokes equations formulated in terms of the stream function, $\psi$, and vorticity, $\omega$. Here, we provide a very brief description of the numerical method applied in solving these Eqs. \eqref{tgVsfe} and \eqref{tgVvte}.


The pseudo-spectral method exploits the power of the discrete Fourier transform to obtain the spatial derivatives with maximum accuracy \cite{arxiv_DNS2021, Deville_etal2002}. In the case of the 2D incompressible Navier-Stokes equations, the method represents the stream function ($\psi$) and vorticity ($\omega$) as discrete sums of complex exponentials in each spatial dimension given by,  

\begin{equation}
    \psi(x, y, t) = \sum_{k_x}\sum_{k_y} \hat{\psi}(k_x, k_y, t) e^{i(k_x x + k_y y)}
\end{equation}

\begin{equation}
    \omega(x, y, t) = \sum_{k_x}\sum_{k_y} \hat{\omega}(k_x, k_y, t) e^{i(k_x x + k_y y)}
\end{equation}

Where \(k_x\) and \(k_y\) are wavenumber components corresponding to the $x$- and $y$-directions, and \(\hat{\psi}\) and \(\hat{\omega}\) are the complex amplitudes in the wavenumber plane. The pseudo-spectral method involves a series of algorithmic steps for spatial discretization, time integration, and finally collating the solution after removing potential sources of the aliasing error in the convective term of Eq. \eqref{tgVvte}.

\begin{itemize}
    \item \textbf{Spatial Differentiation:} The wavenumbers \(k_x\) and \(k_y\) corresponding to the number of grid points are computed. Then the spatial derivative, e.g., $\partial \omega/ \partial x$ is accurately computed by performing the inverse Fourier transform of $\hat{\omega}$ multiplied by $i k_x$, and the second derivative ($\partial^2 \omega/ \partial x^2$) can be obtained by using $-k^2_x$ as the corresponding multiplication factor to obtain $-k^2_x \hat{\omega}$ and then performing the inverse Fourier transform. 
    
    \item \textbf{Solution of Poisson Equation:} Utilizing the computed vorticity \(\omega\) at any given time, the Poisson equation is solved in the spectral domain to determine the stream function \(\psi\) for Eq. \eqref{tgVsfe}. Then $\psi$ is given by the inverse transform of $\hat{\omega}$, multiplied by $(k_x^2 +  k_y^2)^ {-1}$. The spectral amplitude at the origin is taken as zero.  
    
    \item \textbf{Calculation of Velocity Field:} The velocity field \(\vec V\) is obtained by differentiating the stream function \(\psi\) with respect to \(y\) and \(x\), using once again the spectral method for these differentiations.

    \item \textbf{De-aliasing:} A significant challenge associated with the pseudo-spectral method is the complete removal of the aliasing error, which arises due to the truncation of high-wavenumber modes due to finite resolution for all the product term evaluation, which can lead to inaccurate solutions and even instability \cite{HACM, FCFD}. To address this issue, de-aliasing technique of zero-padding is employed. This technique involves introducing additional high-wavenumber modes via zero-padding to accurately capture nonlinear interactions. In the present work, the computational domain is extended from $\pm k_{max}$ to $\pm2 k_{max}$ in both directions for $\vec V$ and $\nabla \omega$ before evaluating the product term in the vorticity transport equations, Eq. \eqref{tgVvte}. The extended zero-padded product used here is different from those advocated in older literature \cite{Canuto_etal} where zero padding is done from $\pm k_{max}$ to $\pm 3/2 k_{max}$. The aliased solution is confined in the extended domain, which is discarded in further computations. 

    \item \textbf{Time Integration:} The four-stage, fourth-order Runge-Kutta (RK4) method is used to advance the solution in time, which can be found in books \cite{HACM}. 
\end{itemize}

In all the reported computations here for the ($4 \times 4$)-cell configuration, a uniform grid with $256 \times 256$ points have been used for $Re = 2000$. For the RK4 time integration method, a time step of 0.025 is adequate as per global spectral analysis \cite{arxiv_DNS2021}. 

\section{Evolution of vorticity and enstrophy for the 2D TGV problem}

Earlier numerical studies of the TGV problem (both 2D and 3D configurations) were restricted to short time intervals focusing on (i) the early inviscid stage of creation of vorticity gradients (as discussed in Brachet {\it et al.} \cite{Brachet_etal1983}) which is explained earlier for interaction of vortices for vortex rings \cite{Saffman}; (ii) subsequent instability of the 2D TGV configurations \cite{Gau_Hattori, Sengupta_etal2DTGV} explained by linear and nonlinear mechanisms; (iii) transition to turbulence and its decay has been discussed for 2D TGV excited by random forcing in Brachet {\it et al.} \cite{Brachet_etal1988} by pseudo-spectral method. 

The canonical TGV problem is special, as the initial condition given by Taylor and Green \cite{Taylor_Green} provides the double-periodic array of vortices of alternate sign in both directions that decays with time analytically given in Eq. \eqref{eqAn}.
To explain the dynamics better for the evolution of the vorticity field, due to the interactions and instabilities of the flow, it is convenient to follow the rotationality in the domain. The measure of the rotationality is given by the enstrophy, which is nothing but the square of the vorticity. The enstrophy transport equation is developed \cite{ETE_2013} from the general vorticity transport equation given in tensor notation by,

\begin{equation} 
\frac{\partial \omega_i}{\partial t}  + u_j\frac{\partial \omega_i}{\partial x_j}   =  \omega_j\frac{\partial u_i}{\partial x_j} + \frac{1}{Re}\frac{\partial^2 \omega_i}{\partial x_j \partial x_j}
\label{3DVTE}
\end{equation}

The first term on the right-hand side is the vortex stretching term and plays an important role for 3D flows only, as it is absent for 
2D flows. Usual tensor operation notations are followed and the enstrophy is defined as, $\Omega_1 = \omega_i \omega_i$, with repeated index imply summation, as the dot product. Furthermore, taking a dot product of Eq. \eqref{3DVTE} with $\omega_i$, one obtains the enstrophy transport equation as \cite{ETE_2013},

\begin{equation} 
\frac{\partial \Omega_1}{\partial t}  + u_j\frac{\partial \Omega_1}{\partial x_j} -2\omega_i\omega_j\frac{\partial u_i}{\partial x_j}   =   \frac{1}{Re}\frac{\partial^2 \Omega_i}{\partial x_j \partial x_j} - \frac{2}{Re}\left(\frac{\partial \omega_i}{\partial x_j}\right) \left(\frac{\partial \omega_i}{\partial x_j}\right)
\label{ETE3D}
\end{equation}

\begin{figure*}
\centering
\includegraphics[width=1\linewidth]{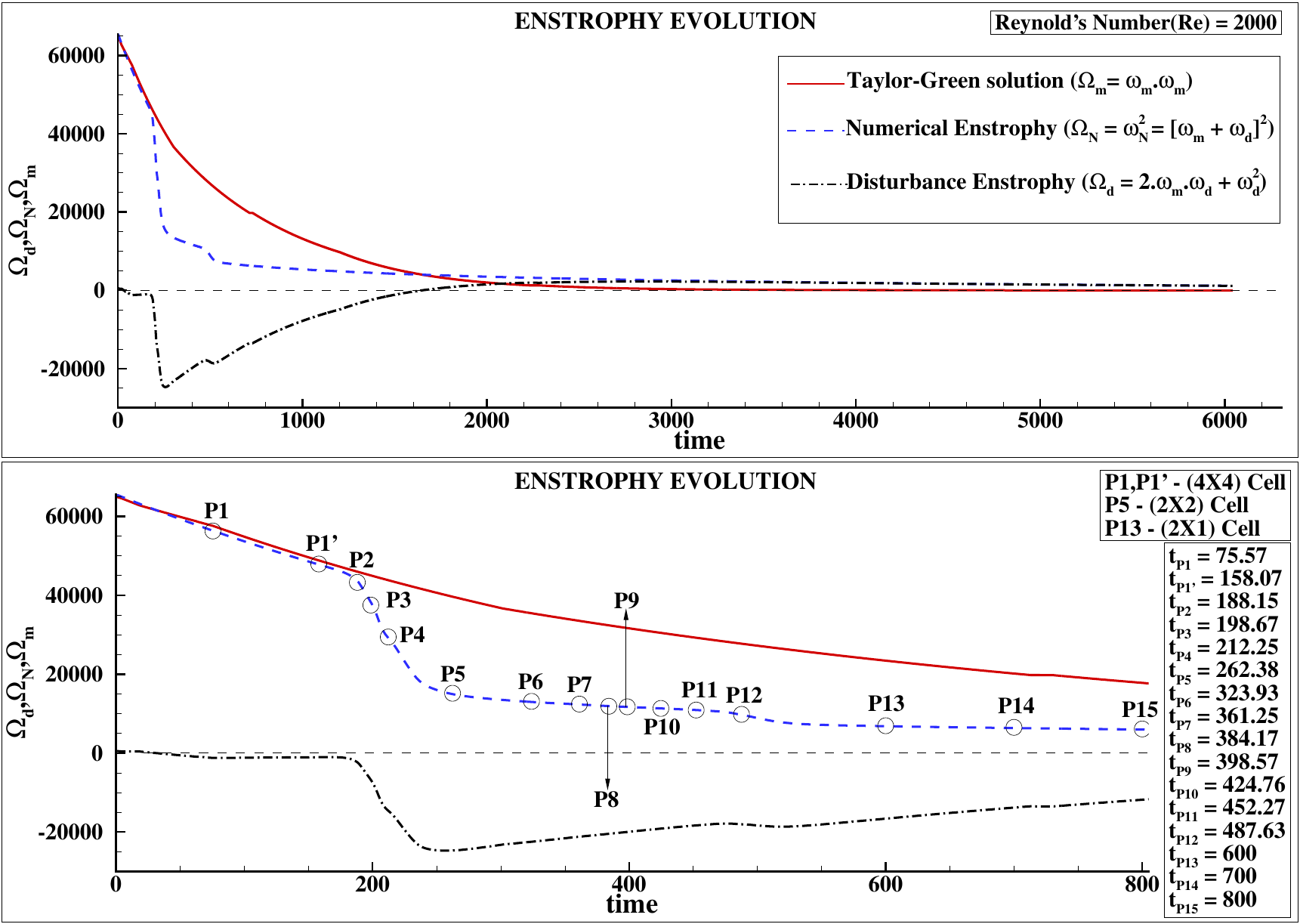}
\caption {Time-evolution of the enstrophy summed over the full domain is plotted up to t = 6000 in the top frame. The bottom frame is the zoomed view of the same up to $t = 800$. The solid (red) line indicates analytical enstrophy, the dotted (blue) line shows numerically computed enstrophy, and the dash-dotted line display the disturbance enstrophy over the full domain.}
\label{Fig2}
\end{figure*}

One notes that the third term on the left-hand side is due to vortex stretching, and is absent for 2D flows. The right-hand side terms arise due to diffusion terms with the first term being diffusive, while the second term with its negative sign is strictly negative. These equations define the point property of the flow and are notationally different from traditional approaches where enstrophy is defined by summing over the full domain \cite{Doerring_Gibbon}. Here, we evaluate the enstrophy for individual points and added for the full domain to show the variation of enstrophy of the domain in figure \ref{Fig2}. In the figure, the analytical solution indicated is for the enstrophy defined as, $\Omega_m (t) = \sum_i \sum_j \omega_m^2 (x_i, y_j, t)$; similarly, the numerically computed total enstrophy is defined as, $\Omega_N(t) = \sum_i \sum_j \omega^2(x_i,y_j,t)$ and the disturbance enstrophy is the difference between the two given by, $\Omega_d (t) = \Omega_N (t) - \Omega_m (t)$. In the top frame of Fig. \ref{Fig2}, these quantities are shown for extended time of up to $t = 6000$. In the bottom frame, the same variations are zoomed during the time up to $t= 800$, with specifically sixteen time instants marked which will help explain the vorticity dynamics. 

The solid line in Fig. \ref{Fig2}, indicates the variation of the analytical equilibrium solution of Taylor-Green for the laminar flow ($\Omega_m (t)$). However, the dashed line in the figure indicates $\Omega_N (t)$ variation obtained by solving Eq. \eqref{tgVvte} by the pseudo-spectral method. The milestones noted in this time series for $\Omega_N (t)$, with the time instants shown in the box indicate important dynamical transition of the vorticity dynamics and associated flow topology. One also notes that the time variation of $\Omega_N(t)$ does not reach the analytical value asymptotically even at $t =6000$, indicating qualitative difference of flow topology. For example, it is noted that there is the flow with ($4 \times 4$) vortical-cells at the times indicated by P1 and P1'. This is followed by the primary instability, which has been satisfactorily explained \cite{Sengupta_etal2DTGV} with the help of disturbance mechanical energy and disturbance enstrophy transport equation. While the detailed flow topology will be provided shortly, the completion of the primary instability will transform the flow from the ($4 \times 4$) vortical-cells to ($2 \times 2$) vortical-cells at P5. Similarly, we just note in passing that at P13 onwards, there will be another transformation to migrate to ($2 \times 1$) vortical-cells that will pervade till the end of the computed flow field at $t =6000$. The time variation of $\Omega_d (t)$ is self-explanatory, i.e. it starts from zero and attains negative value up to about $t \approx 1650$ and thereafter both $\Omega_N(t)$ and $\Omega_d(t)$ remains above $\Omega_m (t)$. 

\begin{figure*}
\centering
\includegraphics[width=1\linewidth]{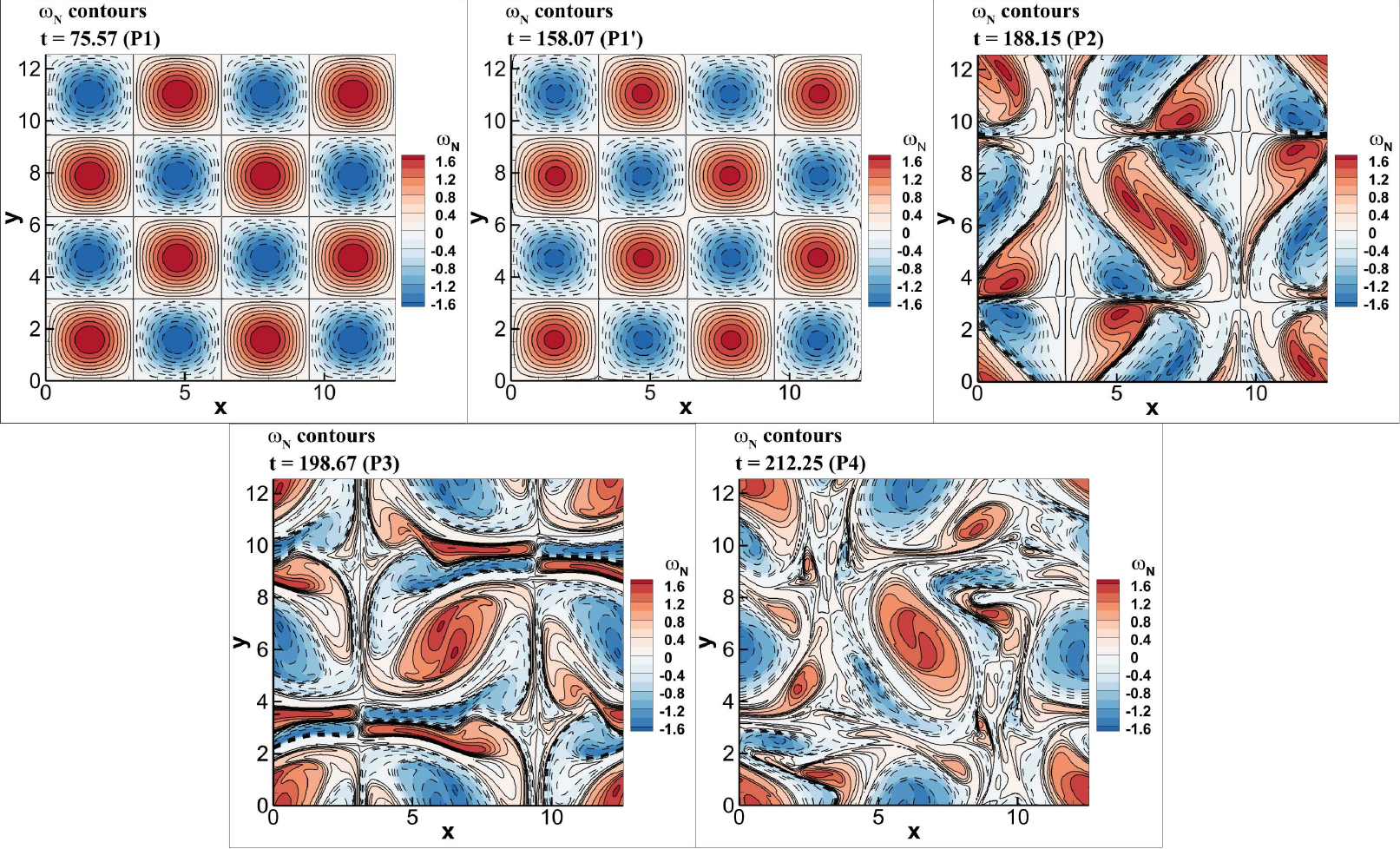}
\caption  {Numerical vorticity contours at indicated times P1, P1', P2, P3 and P4 are shown. Corresponding time instances are noted in the parentheses to display the first bifurcation in its evolution from a clear ($4 \times 4$)-vortical cells to a strongly perturbed ensemble of large number of vortical pairs and stripped vortices are noted; caused by an instability and leading to ($2 \times 2$)-vortical cells.}
\label{Fig3}
\end{figure*}

\begin{figure*}
\centering
\includegraphics[width=1\linewidth]{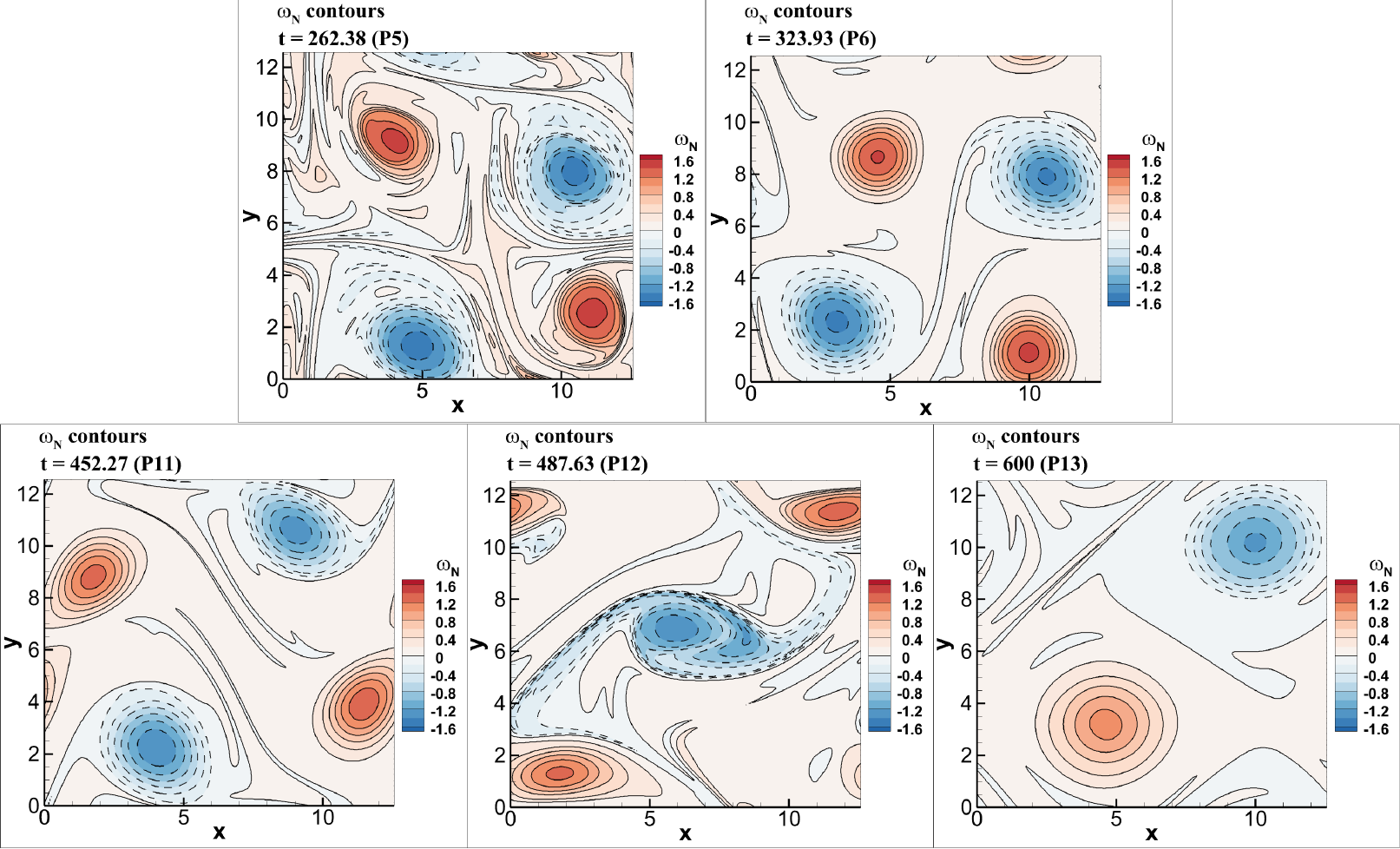}
\caption  {Numerical vorticity contours at indicated times P5, P6, P11, P12 and P13 are shown. Corresponding time instances are noted in the parentheses to display the evolution of second bifurcation from the ($2 \times 2$)-vortical cells to a topological configuration with ($2 \times 1$)-vortical cells is noted; caused by vortex interactions and reconnections.}
\label{Fig4}
\end{figure*}

The first visual difference between analytical and numerical vorticity is noted for the point P1' at $t =158.07$, and such differences grow as shown in Figs. \ref{Fig3} to \ref{Fig4} where the numerical vorticity contours are shown at the indicated times. For the point P1', one notices differences visible in the numerical vorticity contours for $y =0$, $2\pi$ and $4\pi$, with maximum deviations noted at the full saddle points. Various vortices start interacting with others, forming stretched pairs of vortices of the same and opposite signs at subsequent times; with coherent pairs noted at P2 ($t = 188.15$) and an anti-symmetric vortical clusters at P3 ($t= 198.67$). At these early times for $t \leq 160$, the disturbance vorticity magnitudes are very small, and therefore one will not be able to correlate the disturbance vorticity with the computed vorticity at these times. The coherence noted at P3 is disturbed strongly at P4 ($t = 212.25$), and one notices clearly a vortex dipole at the center of the computational domain, and a large numbers of stripped vortices without distinct pattern at P4.

In Fig. \ref{Fig4}, the displayed computed vorticity contours are shown for P5 ($t = 262.38$), P6 ($t = 323.93$), P11 ($t = 452.27$), 
P12 ($t = 487.63$) and P13 ($t = 600$). These times are when one notices primarily ($2 \times 2$)-vortical cells in its evolution to another bifurcation. Thus, in these two figures (Figs. \ref{Fig3} and \ref{Fig4}) one notices two bifurcations with the onset time at P1 when the ($4 \times 4$)-vortical cells are destabilized by the primary instability. This has been described with the help of DETE by Sengupta {\it et al.} \cite{Sengupta_etal2DTGV}. However, a clearer topological transformation to the ($2 \times 2$)-vortical cells shown clearly at P5 in Fig. \ref{Fig4} has not been shown before. In this figure, the second bifurcation to ($2 \times 1$)-vortical cells is apparent with a clearer appearance of it noted at P13.

\begin{figure*}
\centering
\includegraphics[width=1\linewidth]{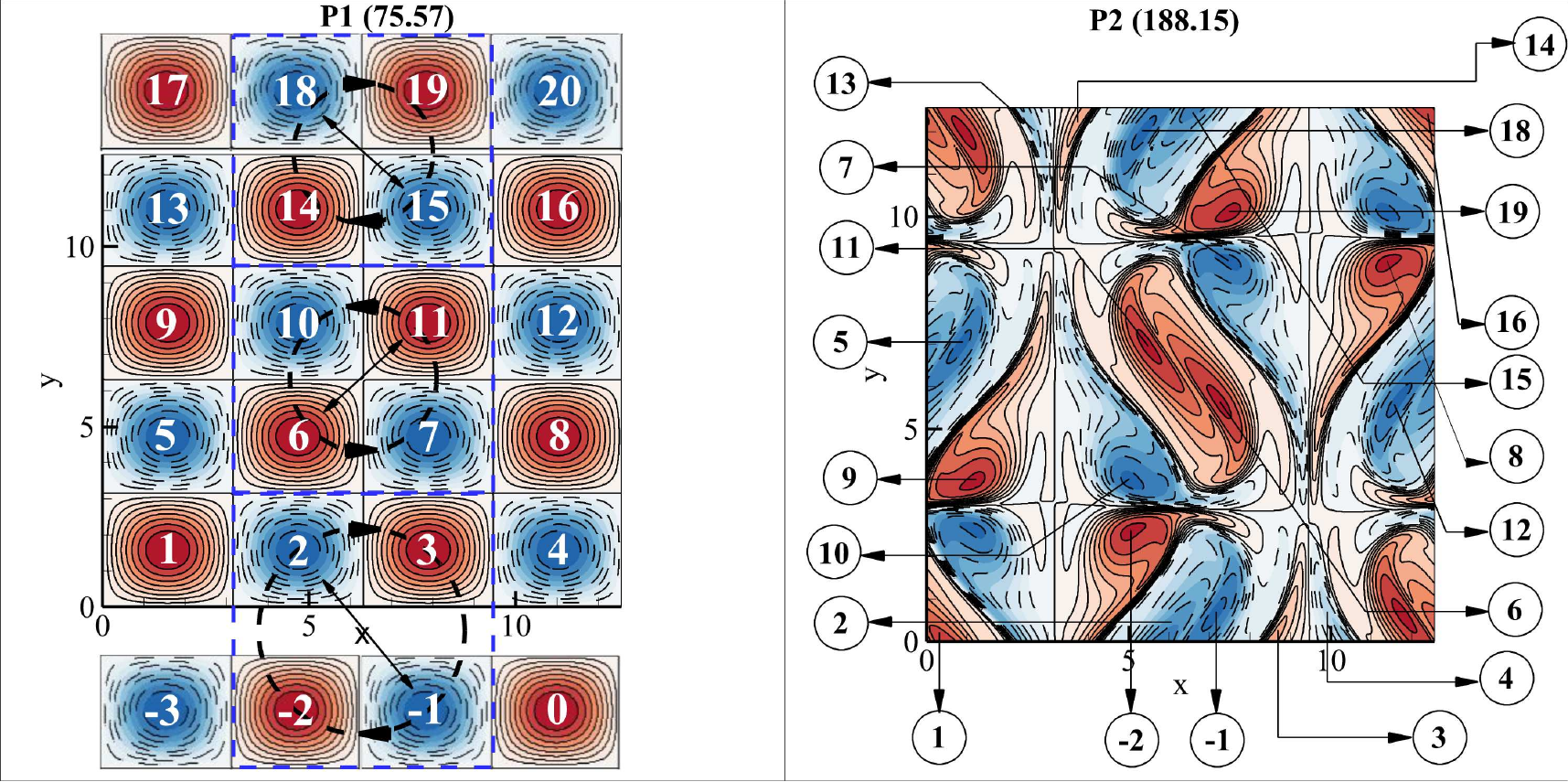}
\caption  {Numerical vorticity contours at P2 are shown here schematically by tracing the interactions of the original ($4 \times 4$)-vortical cells with the cells from the immediate neighbours of the periodic ensemble at P1 ($t = 75.57$) shown on the left. Note the two layers of vortical cells numbered by (-3,-2,-1,0) at the bottom and (17,18,19,20) at the top. The corresponding time instances are mentioned in parentheses. On the right of the figure, the vortex interactions prevalent at P2 ($t= 188.15$) among the plotted vorticity contours are depicted. Of specific interest, is the formation of the vortex-dipole in the center by the elements 6 and 11.} 
\label{Fig5}
\end{figure*}

In Fig. \ref{Fig5}, the numerical vorticity contours are shown for the times ($t= 75.57$ and $188.15$) for P1 and P2 to explain the vortical interactions during the primary instability for the applied periodicity in this multi-cellular configuration. During P1' and afterwards, the vortices interact strongly in the form of stretched vortical elements. Keeping our gaze at the centre of the computation domain, one notices that the positive vortices numbered 6 and 11 approach each other while gyrating in the anti-clockwise direction. Such a vortex dipole formation by positive vortices, causes the negative vortices 7 and 10 to be repelled. This composite picture is drawn here by tracking the 24 vortical cells shown in the left hand side of Fig. \ref{Fig5} and tracking those frame by frame at closely spaced time intervals. At this time, one can also notice the formation of vortex doublets of opposite signs, whose previous locations are marked in the figure. For example, vortices numbered -2 and 10 form one such pair. Similarly, 7 and 19 forms another vortex-dipole of opposite signs, as noted in Fig. \ref{Fig5}. 

This composite picture is drawn here by tracking the 24 vortical cells shown in the left hand side of Fig. \ref{Fig5}, and tracking those, frame by frame at closely spaced time intervals. At this time, one can also notice the formation of vortex doublets of opposite signs, whose previous locations are marked in the figure. For example, vortices numbered -2 and 10 form one such pair. Similarly, 7 and 19 forms another vortex-dipole of opposite signs, as noted in Fig. \ref{Fig5}. 

Also, one sees in Fig. \ref{Fig2}, the increased coherence between the numerical and disturbance vorticity contours. In the subsequent frame for P3 at $t= 198.67$ in Fig. \ref{Fig3}, one can note the compaction of the approaching vortices 6 and 11 further, while the vortex doublets of opposite signs become stretched more in the horizontal directions. The other two vortex doublets of negative signs also become more compact and keep rotating together. At $t=212.25$, one can see the vortex doublets of the same sign in the center column of the frame. During this phase, one can also notice a significant drop of the enstrophy shown in Fig. \ref{Fig2}, all the way up to P5 ($t=250$). Since, the enstrophy for flows is strongly related to dissipation, such drastic loss of enstrophy implies intensification of vorticity gradient of the vortex dipoles of opposite signs destroying each other, as has been suggested also by Saffman \cite{Saffman}. Here, the plotted vortical contours indicate that during this first bifurcation stage, the vortex-dipoles (6 and 11); (-1 and 2); (15 and 18) and a positive vortex-dipole which is noted at the top left and at the bottom right; survive in the computational domain at P5 as coherent vortices.

The transformation of the original ($4 \times 4$)-cells to ($2 \times 2$)-cells indicates migration from one equilibrium state to another, and so is the subsequent transformation from the ($2 \times 2$)-cells to ($2 \times 1$)-cells. The merger of cells is indicative of vortex connection of the TGV, which has not been reported before. Brachet {\it et al.} \cite{Brachet_etal1983}have attributed this to the viscous nature, which prevents the formation of inviscid singularities. In contrast, this "may allow inviscidly formed structures to coalesce into larger ones; and may also induce new instabilities" during the transient stages of decaying turbulence. The sequence of events from ($4 \times 4$)-cell to this final ($2 \times 1$)-cell configuration are associated with coherent vortices, and thus one should note this as a vortex-merger, rather than inverse cascade with 2D turbulence. In Fig. \ref{Fig1}, the computed asymptotic state of the TGV flow field is shown in terms of the vorticity contours. In Fig. \ref{Fig2}, one can notice that the computed flow field does not approach even at $t= 6000$, the analytical solution due to Taylor-Green \cite{Taylor_Green}, which achieves a nearly vanishing value for the vorticity with ($4 \times 4$)-cells. Instead one notices a coherent ($2 \times 1$)-cells with a circular positive vortex and a negative vortex that resembles a rhombus, with no other vortical structures seen in the computational domain.  

\section{Vorticity dynamics: Enstrophy transport for 2D TGV}

The vorticity dynamics of 2D TGV needs interpretation with respect to enstrophy transport for 2D flows \cite{Doerring_Gibbon}, as explained \cite{ETE_2013} that the enstrophy transport equation for 3D flows transforms to the following equation for 2D flow as,

\begin{equation}
\frac{D\Omega_1}{Dt} = \frac{2}{Re} \left[\frac{1}{2}\nabla^2\Omega_1 - \left(\nabla\omega\right)^2\right]
\label{ETE2D}
\end{equation}



It has been reasoned by researchers \cite{YDS, Kerr} that enstrophy and pressure statistics in turbulent flows are noted in the dissipation experienced at high Reynolds numbers. Specifically for 2D flows (in the absence of vortex stretching), the enstrophy transport equation given in Eq. \eqref{ETE2D} has the first term on the right-hand side due to diffusion, while the second term is strictly positive definite indicating its effect as a dissipation term. It has also been stated \cite{Doerring_Gibbon} that for a strictly periodic flow, the diffusion term integrated over the whole domain goes to zero, without any contribution. Thus, for strictly periodic 2D flows, the enstrophy transport reduces to $\frac{D\Omega_1}{Dt} = - \left(\nabla\omega\right)^2$ or $\Omega_1$ approaches zero asymptotically. This is obeyed by the analytical Taylor-Green solution of the TGV problem with vanishing enstrophy, but topologically retaining the original ($4 \times 4$)-cells of Fig. \ref{Fig1} due to the exponential decay of the solution with time. DNS of 2D TGV shows that the time dependence is not an exponentially decaying function obtained by the pseudo-spectral method \cite{arxiv_DNS2021}. Instead, the computed flow resembles the wake vortices noted behind aircraft where Crow instability \cite{Crow} also creates long enduring vortex rings, as discussed by Saffman \cite{Saffman_book}. 


The episodic description of the evolving vorticity field in the previous section alludes to certain physical roles played by vortex interactions \cite{Mariotti_etal}, the role of pressure gradient in vorticity dynamics \cite{Brachet_etal1983, Brachet1991}, and the behaviour of decaying turbulence \cite{Brachet_etal1988}. There are also accounts of developing singularity in TGV problem solved by an inviscid approach \cite{MOF, Brachet_etal1983}.

\subsection{Role of pressure gradient for TGV problem}

The interactions between vorticity distributions separated at a distance has attracted researchers' attention in the study of flow instability. Failure to create a transition for wall-bounded shear layer by free stream acoustic excitation \cite{Schubauer_Skramstad} has spawned the subject of receptivity \cite{Sengupta21, Morkovin}, and led to the coining of the term, shear sheltering \cite{Hunt_Durbin} for free stream vortical excitation on zero pressure gradient boundary layer. However, Morkovin \cite{Morkovin} proposed that such long-distance interaction causes unsteady static pressure that can provide the seed for transition. In the context of TGV problem, Brachet \cite{Brachet1991} wrote down the Poisson equation for static pressure as,

\begin{equation}
    \frac{2}{\rho} \nabla^2 p = \Omega_1- \sigma^2
\label{SPPE}
\end{equation}

\noindent where the last term on the right-hand side is due to local energy dissipation given by, $\rho \nu \sigma^2$, with $\sigma^2 = \frac{1}{2} \sum_{ij} (\partial_i v_j + \partial_j v_i)^2$. This led to the conjecture that the enstrophy is correlated with low pressure, while the symmetric part of the rate of strain tensor is related to zones of high pressure. Brachet \cite{Brachet1991} also observed that many theories of intermittency "do not take into account the difference of exponent that we observe ... between energy dissipation and square vorticity". This indicates that viewing the static pressure or the kinetic energy is not adequate.

A major change of perspective took place in the study of receptivity and transition, with the point of view that instead of looking at only the kinetic energy, one should look at total mechanical energy\cite{VII_JFM03}. Interested readers are urged to look at the developed disturbance mechanical energy (E) equation for many applications in Sengupta \cite{Sengupta12, Sengupta21}. 

The Poisson equation for the distribution of $E$ is shown to be driven by the enstrophy on the right-hand side \cite{Sengupta12, Sengupta21}. Qualitatively, the sign of the right-hand side indicates the presence of source and sink of $E$, with negative and positive signs, respectively. In the previous study of 2D TGV with ($2 \times 2$)-cells \cite{Sengupta_etal2DTGV}, the concept of DETE was introduced, which traces the rotationality in the domain of interest. This is described in the following.

\begin{figure*}
\centering
\includegraphics[width=1\linewidth]{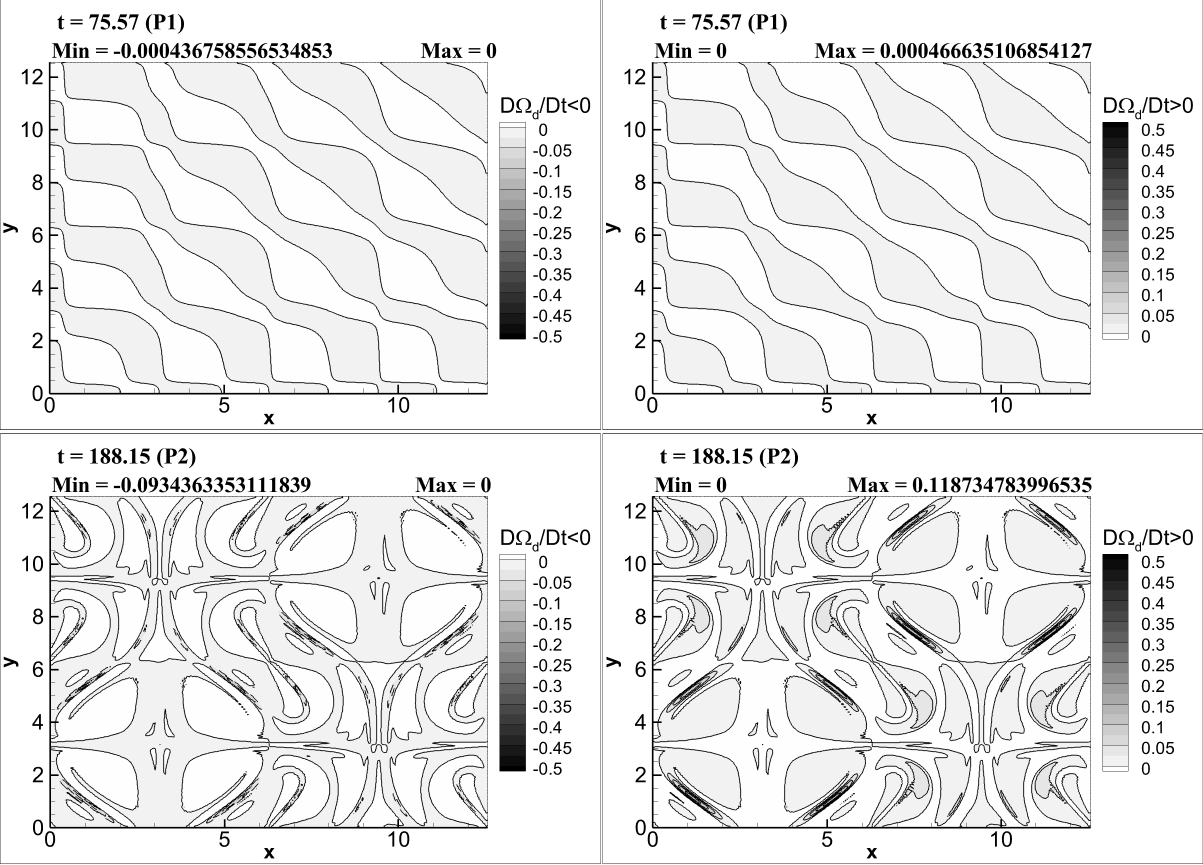}
\caption  {The nonlinear growth rate contours of $\Omega_d$ for the TGV problem at the indicated time instants: $t= 75.57$ (top) and $t= 188.15$ (bottom). The condition of growth corresponds to $\frac{D\Omega_d}{Dt} > 0$ for $\Omega_d > 0$ (right) and $\frac{D\Omega_d}{Dt} < 0$ for $\Omega_d < 0$ (left).}
\label{Fig6}
\end{figure*}

\subsection{Disturbance enstrophy transport equation: Application to the TGV problem}

In developing instability theory, the dynamical system, in an equilibrium state, is studied for its receptivity to omnipresent background disturbances or deterministic imposed disturbance \cite{Sengupta21}. In that context, the 2D TGV problem with the analytical solution for the spatially periodic domain (with analytic enstrophy, $\Omega_m$) is studied, and the evolving enstrophy is indicated in Fig. \ref{Fig2}. If one represents $\Omega_1$ as a sum of equilibrium and disturbance components: $\Omega_1 = \Omega_m + \epsilon \Omega_d$, along with the primary variables given by, $\omega = \omega_m + \epsilon_1 \omega_d$ and $\vec{V} = \vec{V}_m + \epsilon_1 \vec{V}_d$. Then the growth/ decay rate of enstrophy can be written for 2D flows as \cite{Asengupta_etal2018},

\begin{equation}
 \frac{D\Omega_d}{Dt}= \frac{1}{Re} \frac{\partial^2\Omega_d}{\partial x_i \partial x_j} - \frac{2}{Re} \left(\frac{\partial \omega_{im}}{\partial x_j} \right) \left(\frac{\partial \omega_{id}}{\partial x_j} \right)
\label{DETE}
\end{equation}

Here, $\Omega_1$ is positive definitive, but $\Omega_d =2 \omega_m \cdot \omega_d$ can be either positive or negative. Thus, the conditions of instability correspond to $\frac{D\Omega_d}{Dt} >0$ for $\Omega_d >0$ and $\frac{D\Omega_d}{Dt} <0$ for $\Omega_d <0$. The instability is determined as to how the mean vorticity interacts with the disturbance vorticity, as determined by the numerical evaluation of the right hand side of Eq. \eqref{DETE}. 

\begin{figure*}
\centering
\includegraphics[width=1\linewidth]{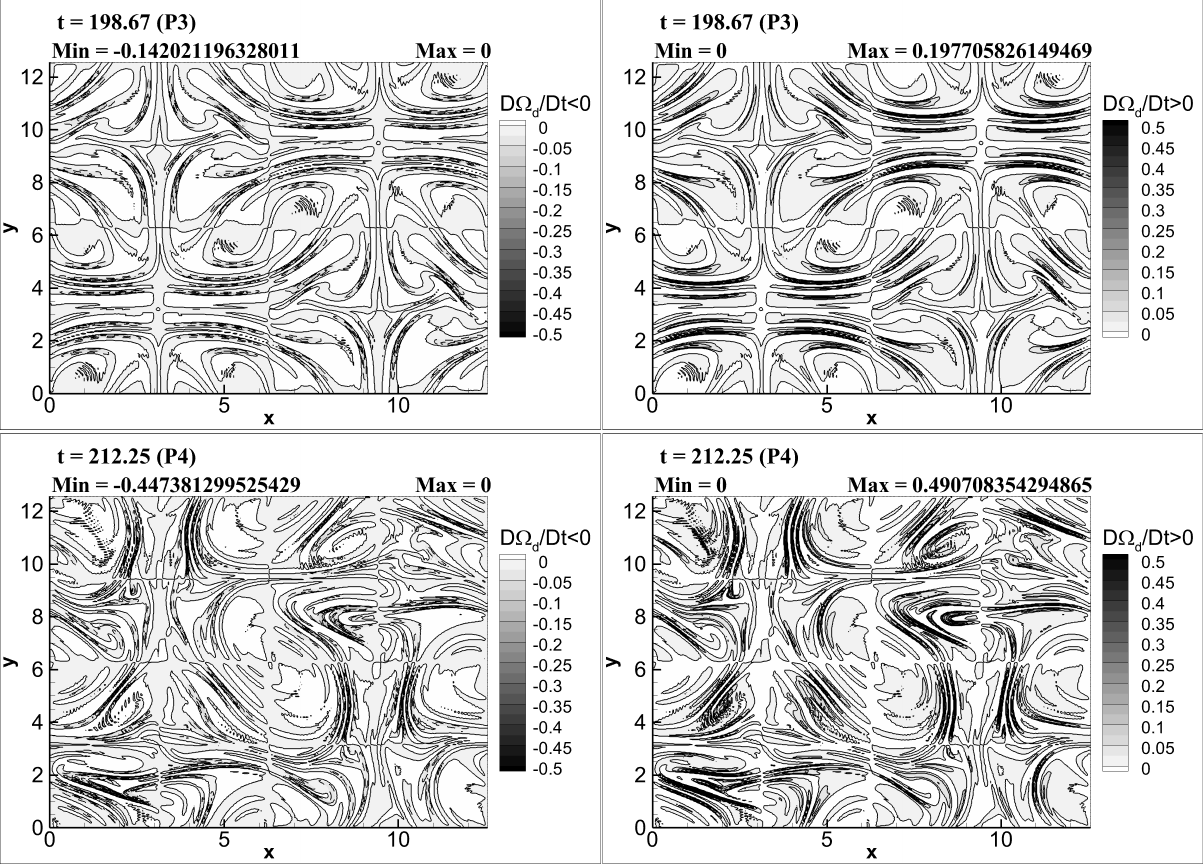}
\caption  {The nonlinear growth rate contours of $\Omega_d$ for the TGV problem at the indicated time instants: $t= 198.67$ (top) and $t= 212.25$ (bottom). The condition of growth corresponds to $\frac{D\Omega_d}{Dt} > 0$ for $\Omega_d > 0$ (right) and $\frac{D\Omega_d}{Dt} < 0$ for $\Omega_d < 0$ (left).}
\label{Fig7}
\end{figure*}

\begin{figure*}
\centering
\includegraphics[width=1\linewidth]{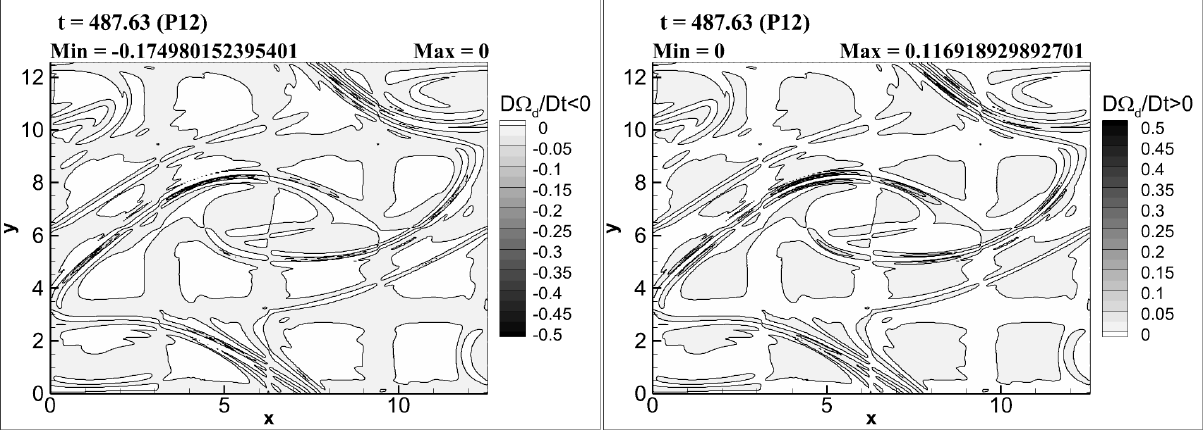}
\caption  {The nonlinear growth rate contours of $\Omega_d$ for the TGV problem at the indicated time instants: $t= 487.63$. The condition of growth corresponds to $\frac{D\Omega_d}{Dt} > 0$ for $\Omega_d > 0$ (right) and $\frac{D\Omega_d}{Dt} < 0$ for $\Omega_d < 0$ (left).}
\label{Fig8}
\end{figure*}

In Figs. \ref{Fig6} to \ref{Fig8}, both these conditions for the growth of $\Omega_d$ are shown at indicated times. It is evident that before the onset of primary instability, there was hardly any growth noted anywhere in the domain for the point P1 ($t= 75.57$) in Fig. \ref{Fig6}. While the enstrophy is noted to increase at $t= 188.15$ for both the signs, the growth is seen to intensify in Fig. \ref{Fig7} at $t= 212.25$, where the interactions are seen to reach peak values. Thereafter, the enstrophy growth rate is seen to come down and is not shown further, except the event noted in Fig. \ref{Fig8}, where one can note the transition from ($2 \times 2$)-cell to ($2 \times 1$)-cell configuration, with the dominant vortex-merger is noted at the center of the domain. 

\begin{figure*}
\centering
\includegraphics[width=1\linewidth]{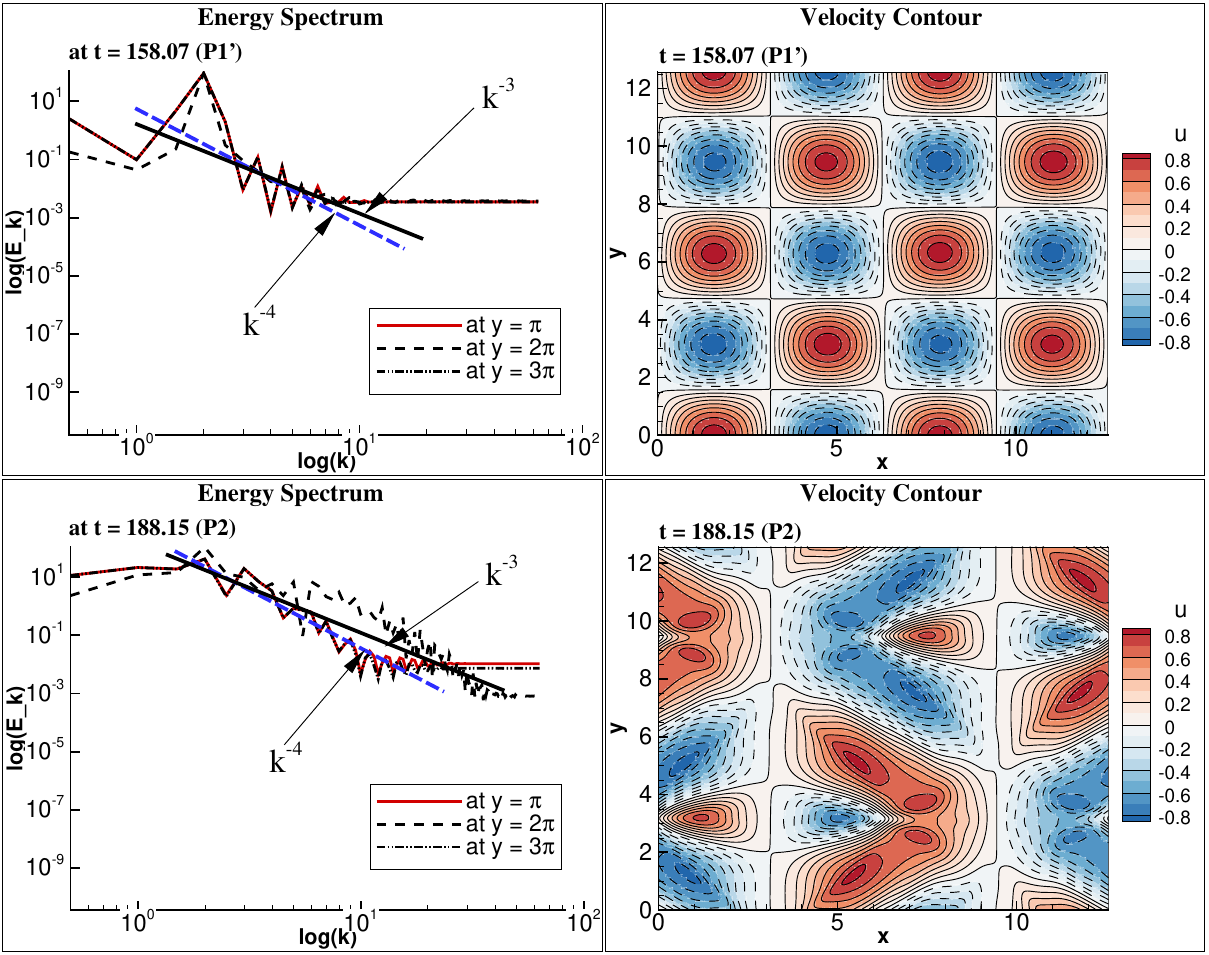}
\caption  {The energy spectrum $(E(k))$ plots for the TGV problem at the indicated time instants: $t= 158.07$ and $t= 188.15$ with variations along $x$-direction, for fixed vertical locations of $y = \pi$, $2\pi$ and $3\pi$ indicated by solid (red) line, dashed (black) line and dashed-dot (black) line respectively are shown on the left-hand side of the frame. The right-hand side of the frame shows the corresponding velocity contours.}
\label{Fig9}
\end{figure*}

\section{Energy spectrum at selected vertical locations: Deterministic dynamics or decaying turbulence?}

In the introduction we raised some queries which motivated us for the present research, by using the pseudo-spectral method, as also used by Brachet {\it et al.} \cite{Brachet_etal1988} for the 2D TGV problem. Instead of using the lower order Runge-Kutta time integration scheme by previous authors, in the present simulations, RK4 time integration scheme is used following the global spectral analysis of the space-time discretization \cite{arxiv_DNS2021}. The authors in the earlier study \cite{Brachet_etal1988} did not use the Taylor-Green analytic solution for the initial condition and used random perturbation to obtain results which has been interpreted as the decaying turbulence (based on an inertial range) from the energy spectrum exponent varying between -4 and -3. The exponent value of -4 has been identified by the authors to be associated with isolated vorticity gradient sheet, as postulated also by Saffman \cite{Saffman}. If the dynamics is that of purely
2D turbulence, then the energy spectrum would vary as $k^{-3}$, according to a developed theory \cite{Kraichnan, Batchelor}. More detailed analysis of spectrum and associated enstrophy cascade have been provided by other authors \cite{Doerring_Gibbon, ETE_2013}.  
In the following, we try to provide results to indicate if the 2D decaying turbulence \cite{Brachet_etal1988} is due to random excitation for the pseudo-spectral method used with different initial condition and different domain size. The energy spectrum obtained following the deterministic dynamics of 2D Navier-Stokes equation is presented next. 

\begin{figure*}
\centering
\includegraphics[width=1\linewidth]{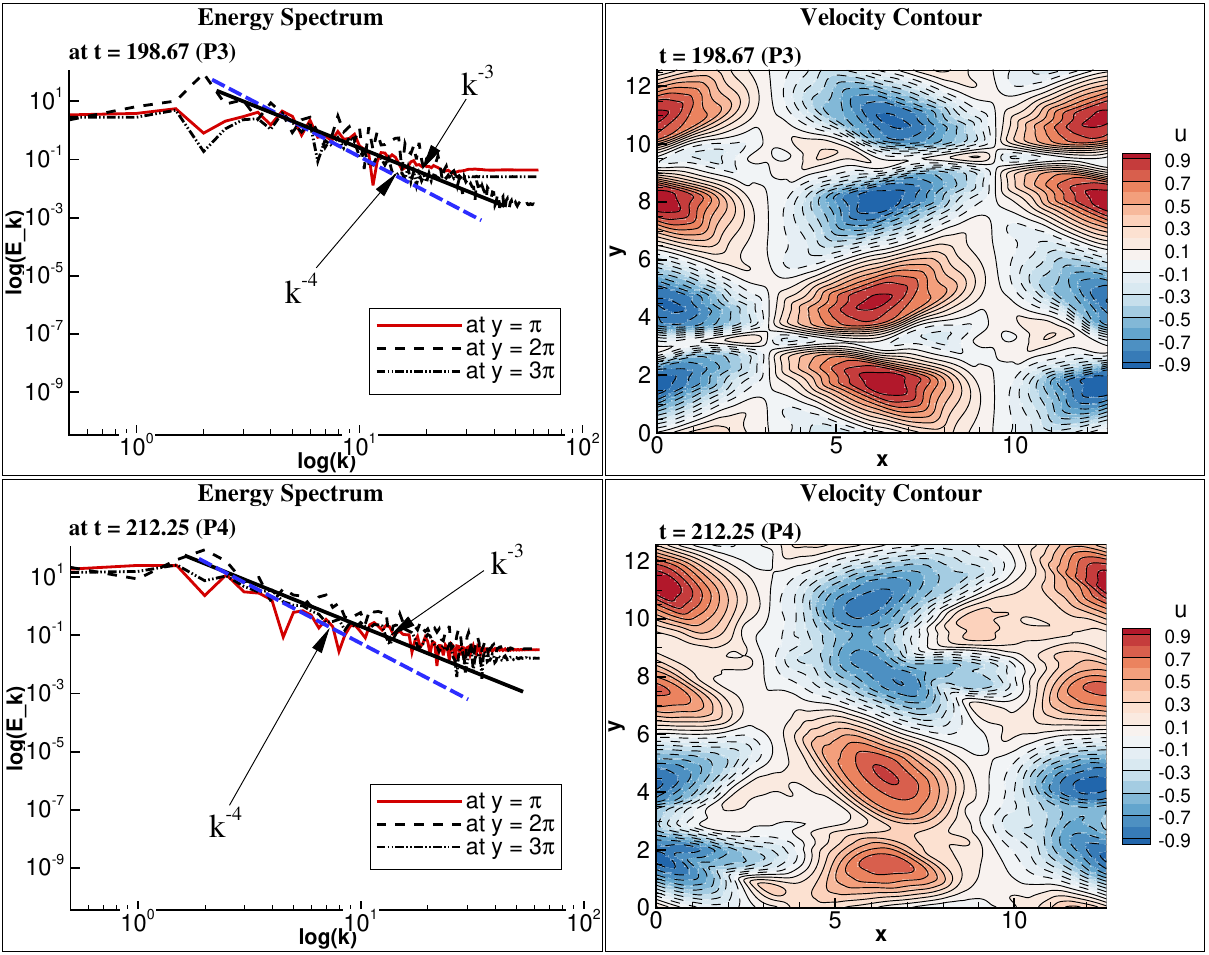}
\caption  {The energy spectrum $(E(k))$ plots for the TGV problem at the indicated time instants: $t= 198.67$ and $t= 212.25$ with variations along $x$-direction, for fixed vertical locations of $y = \pi$, $2\pi$ and $3\pi$ indicated by solid (red) line, dashed (black) line and dashed-dot (black) line respectively are shown on the left-hand side of the frame. The right-hand side of the frame shows the corresponding velocity contours. }
\label{Fig10}
\end{figure*}

\begin{figure*}
\centering
\includegraphics[width=1\linewidth]{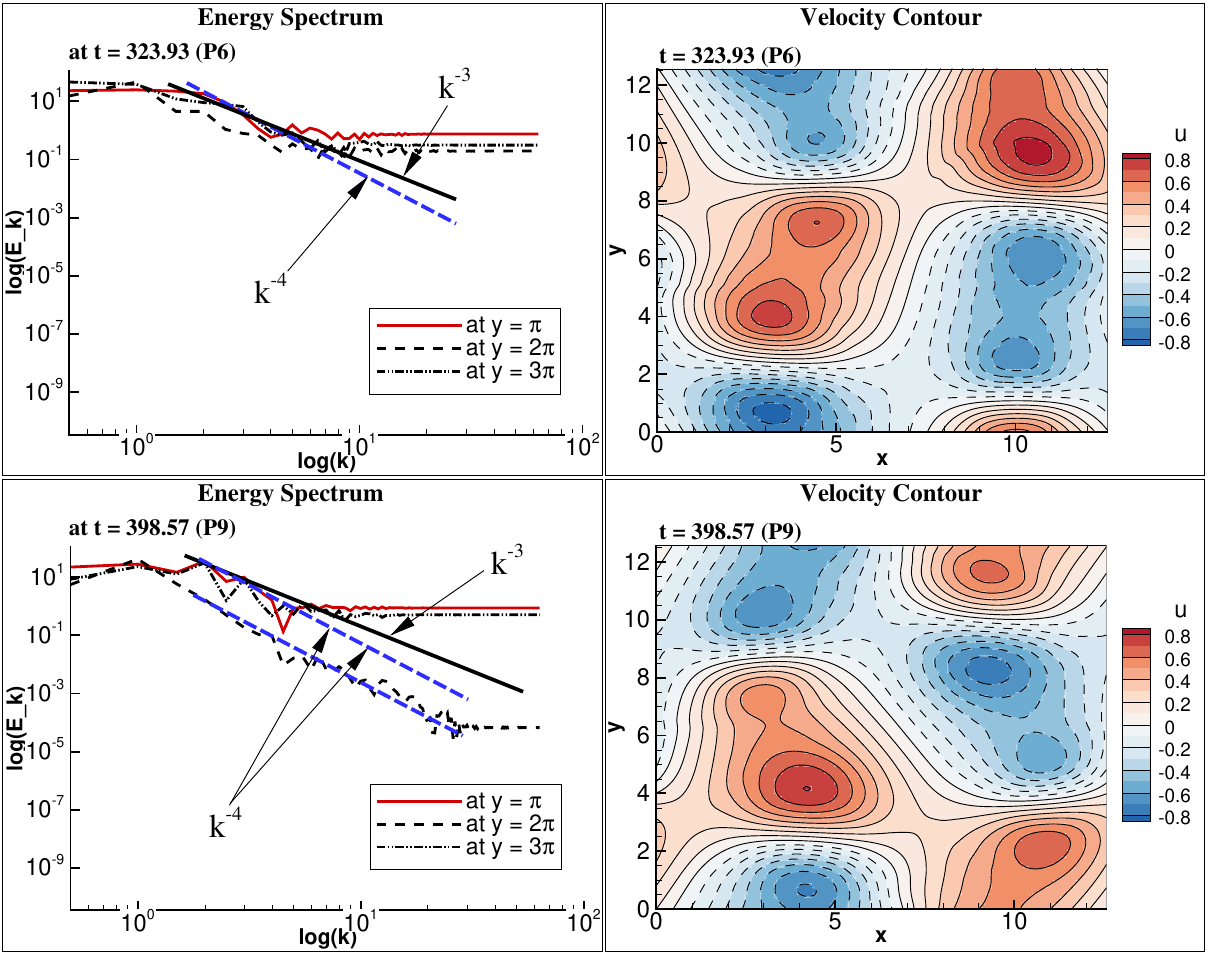}
\caption  {The energy spectrum $(E(k))$ plots for the TGV problem at the indicated time instants: $t= 323.93$ and $t= 398.57$ with variations along $x$-direction, for fixed vertical locations of $y = \pi$, $2\pi$ and $3\pi$ indicated by solid (red) line, dashed (black) line and dashed-dot (black) line respectively are shown on the left-hand side of the frame. The right-hand side of the frame shows the corresponding velocity contours. }
\label{Fig11}
\end{figure*}

\begin{figure*}
\centering
\includegraphics[width=1\linewidth]{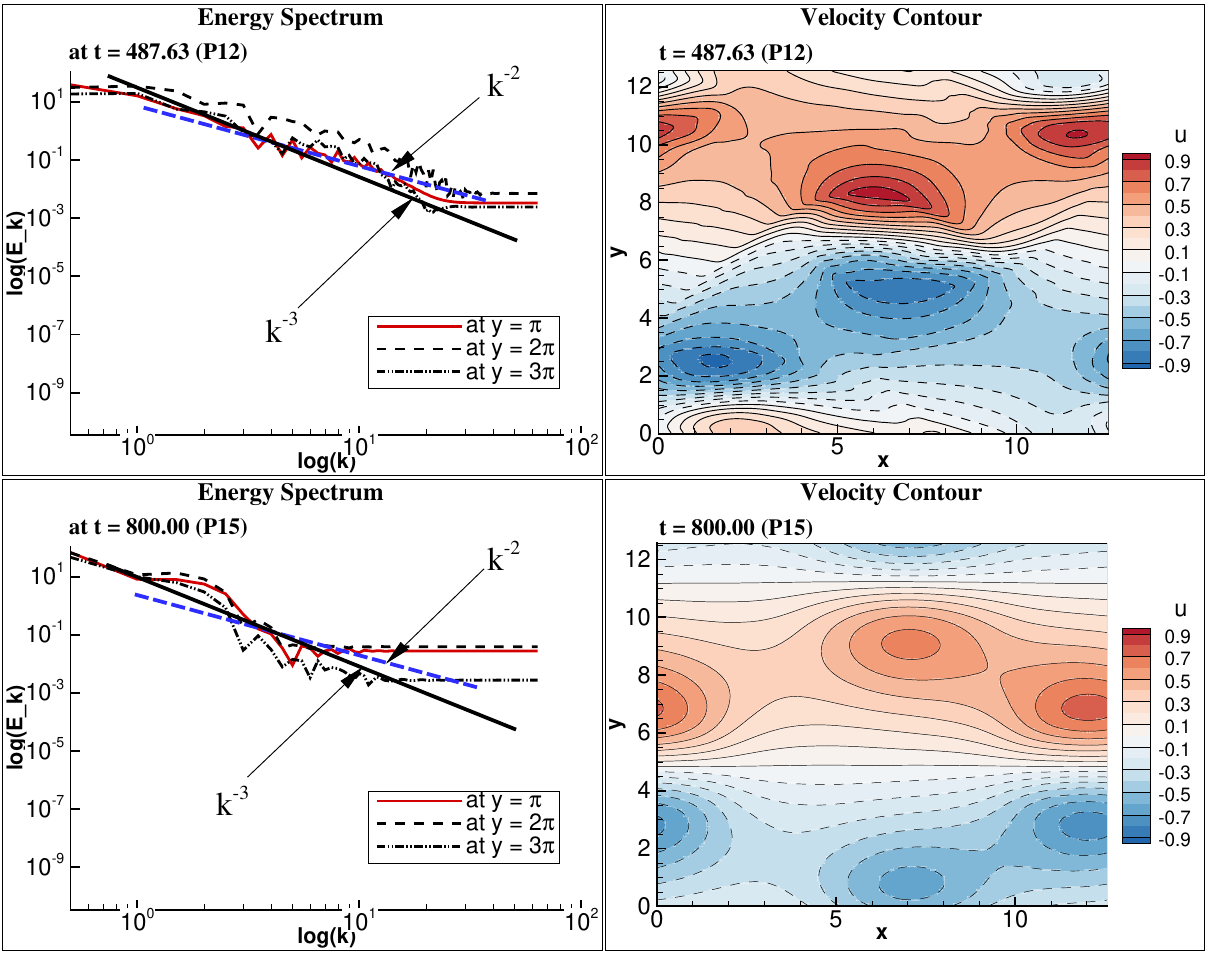}
\caption  {The energy spectrum $(E(k))$ plots for the TGV problem at the indicated time instants: $t= 487.63$ and $t= 800.00$ with variations along $x$-direction, for fixed vertical locations of $y = \pi$, $2\pi$ and $3\pi$ indicated by solid (red) line, dashed (black) line and dashed-dot (black) line respectively are shown on the left-hand side of the frame. The right-hand side of the frame shows the corresponding velocity contours. }
\label{Fig12}
\end{figure*}

In Figs. \ref{Fig9} to \ref{Fig12}, the energy spectrum $(E(k))$ of 2D TGV is shown at specific times for variations along $x$-direction, for fixed vertical locations of $y = \pi$, $2\pi$ and $3\pi$, which are all along the interior of the domain. At $t =0$, these three lines are aligned with the full-saddle points in the horizontal direction with zero vorticity. However, after the onset of the primary instability, these will not be along special symmetry lines.

As this 2D TGV problem is doubly periodic, the $x$-component of velocity contours are also shown at each time, as any of the velocity component will be equally contributing to the energy spectrum. 

In Figs. \ref{Fig9} and \ref{Fig10}, such variations are shown for times during the primary instability at P1'($t=158.07$), P2 ($t=188.15$), P3 ($t=198.67$) and P4 ($t=212.25$). During this phase, one observes discrete vortices suffering vortex stripping and merger with only finite number of vortices and the flow field hardly corresponds to the 2D turbulent flow field \cite{Nastrom_Gage, Sengupta_etalPRE2012}. The earlier energy spectrum results of 2D TGV problem solved when excited by random forcing \cite{Brachet_etal1988} showed the exponent to lie between -4 and -3, and for the sake of comparison, in all these figures showing the energy spectrum, two lines have been drawn with the slope corresponding to these two values. At $t=158.07$ (P1'), $E(k)$ distribution in Fig. \ref{Fig9} displays the peak corresponding to the length scale of the $u$-velocity contours shown on the right hand side. The other peaks are the superharmonics of this fundamental and their alignment with the $k^{-3}$ is mere coincidental, while the $k^{-4}$ line is more meaningful, as was suggested by Saffman \cite{Saffman_book}. Once the primary instability becomes dominant, one notices the appearance of vortex stripping, stretching in the horizontal direction by the vortex sheets of opposite signs, reconnection and merger of vortices of same sign, as given in the vorticity contours in Fig. \ref{Fig5} at $t = 188.15$. The energy spectrum at this instant is shown in the bottom left panel of Fig. \ref{Fig9}, with the corresponding $u$-contours shown in the right panel. As the $u$-velocity distribution for $y= \pi$ and $3\pi$ are identical, $E(k)$ distributions also show for these heights an overlapping variation. The distribution for $y =2\pi$ is distinctly different and displays a high wavenumber range that appears to follow $k^{-4}$ line. However for lower wavenumbers, one can also guess the existence of $k^{-3}$ trend. This only points to the fact that noticing the exponent of energy spectrum alone cannot unambiguously show the existence of 2D turbulence. 
During the primary instability, the multi-polar vortical interactions indicate the simultaneous presence of multiple exponents for $E(k)$.

In Fig. \ref{Fig10}, the energy spectrum (left) and the corresponding $u$-component of velocity (right) are shown at $t= 198.67$ (P3) and $t = 212.25$ (P4), which are also during the primary instability that takes the flow from the ($4 \times 4$)-cells to ($2 \times 2$)-cells via the first bifurcation. While one can note significant vortex stripping during these later stages of the first bifurcation with the appearance of small scale vortices, the $u$-component of velocity distribution maintains significant coherence in Fig. \ref{Fig10} at $t= 198.67$, and lesser coherence at $t = 212.25$. $E(k)$ distribution for $y = \pi$ and $3\pi$ once again shows similar variations, while the distribution for $y =2\pi$ is distinctly different. The present results are for the 2D TGV multi-periodic cases which are reported here for the first time, and the results are partly consistent with previous observations \cite{Brachet_etal1988, Saffman, Saffman_book} for the exponents of $E(k)$. 

In Fig. \ref{Fig4}, we have noted from the displayed vorticity contours the existence of $(2 \times 2)$-vortical cells and $(2 \times 1)$-vortical cells following the two bifurcations. In Figs. \ref{Fig11} and \ref{Fig12}, the $u$-velocity contours (right) and the corresponding $E(k)$ distribution at three selected heights (left) are shown for cases with $(2 \times 2)$-vortical cells and $(2 \times 1)$-vortical cells. The $u$-velocity cntours in Figs. \ref{Fig11} and \ref{Fig12} display regular contours to begin with, while at $t = 800$ (P15), for the $(2 \times 1)$-cells case, one notices alternate streaks of negative and positive velocity values with recirculating structures within. For $t =323.93$ (P6) all the three $E(k)$ distributions are similarly aligned with no strong discernible trend with either $k^{-4}$ or $k^{-3}$-like variations. For $t =398.57$ (P9) the $E(k)$ distributions are identical for $y = \pi$ and $3\pi$, while the results for $y=2\pi$ shows significantly quiet energy spectrum, as the induced velocity at this height is smaller by orders of magnitude.

In Fig. \ref{Fig12}, the $u$-velocity contours and $E(k)$ spectrum are shown at later times ($t= 487.63$ and 800) where these are characterized by the second bifurcation from $(2 \times 2)$ to $(2 \times 1)$-cells. The energy spectrum shows an atypical variation of $k^{-2}$ variation for P12. For this time of $t =487.63$, one notices the merger of the negative vortices in the center of the domain, with severely distorted $u$-velocity contours. Following this bifurcation, the spectrum and the $u$-velocity contours take a very deterministic dynamics with streak formation in the velocity contours. This implies that in the presence of only two vortices in the computational domain, the energy spectrum will be significantly height dependent and the center of the domain will show a depleted energy spectrum.

\section{Summary and conclusion}

The current study investigates the dynamics of the 2D Taylor-Green vortex through DNS of the incompressible Navier-Stokes equation. This study uses the $(\psi,\omega)$-formulation on a uniform grid with ($256 \times 256$) grid points in the domain $0 \le (x, y) \le 4\pi$, with sixteen vortical cells, as depicted in Fig. \ref{Fig1}. Pseudo–spectral method is used for the spatial discretization by Fourier series, and a four-stage, fourth-order Runge–Kutta scheme is used for time integration.
Here, the study is performed by integrating the Navier-stokes equation for an extended period of time for the first time in multiple periods of TGV cells. This is done to derive insights on vortex interactions that lead the flow back to an ordered minimal configuration. 
The flow evolution is shown by enstrophy versus time for the whole domain in Fig. \ref{Fig2}. This helps in identifying instabilities/ bifurcations which is described in detail, with very specific milestones indicated.   
In Fig. \ref{Fig3} (P1 to P4), the primary instability is discussed. This is different from earlier studies with ($2 \times 2$)-cells of TGV to explain primary instability by performing disturbance enstrophy transport equation (DETE) analysis. Sengupta {\it et al.} \cite{Sengupta_etal2DTGV} have shown it only for primary instability and Brachet et al.\cite{Brachet_etal1983} studied the evolution and subsequent decay of turbulence caused by random initial perturbation without using the analytical solution of Taylor and Green. 
In Fig. \ref{Fig4}, we show the long-term evolution of this ($4 \times 4$)-cells into ($2 \times 2$)- and ($2 \times 1$)-cells configurations. This ($2 \times 2$)-cells forms after vortex stripping and merger of original ($4 \times 4$)-cells whose details are specifically given in Fig. \ref{Fig5}.
In Figs. \ref{Fig6} to \ref{Fig8}, the growth of disturbances is explained by the DETE analysis at typical time instants starting from the original ($4 \times 4$)-cells to the creation of ($2 \times 1$)-cells.

In Figs. \ref{Fig9} to \ref{Fig12}, the $u$-velocity contours and $E(k)$ spectrum are shown at different time instants. The energy spectral  distribution during discrete vortex interactions implies that merely the existence of the exponent $k^{-3}$ does not imply 2D turbulence for such multi-cellular configurations of 2D TGV.

\section*{AUTHOR DECLARATIONS}

\subsection*{Conflict of Interest}
The authors have no conflicts to disclose.

\section*{DATA AVAILABILITY}
The data that support the findings of this study are available from the corresponding author upon reasonable request.

\bibliography{TGV}
\end{document}